# Bayesian modelling of statistical region- and family-level clustered ordinal outcome data from Turkey

Özgür Asar

*Department of Biostatistics and Medical Informatics,*
*Acıbadem Mehmet Ali Aydınlar University, 34752*
*İstanbul, Turkey*
*ozgurasarstat@gmail.com*
*orcid.org/0000-0003-0603-1409*

**Abstract**

This study is concerned with the analysis of three-level ordinal outcome data with polytomous logistic regression in the presence of random-effects. It is assumed that the random-effects follow a Bridge distribution for the logit link, which allows one to obtain marginal interpretations of the regression coefficients. The data are obtained from the Turkish Income and Living Conditions Study, where the outcome variable is self-rated health (SRH), which is ordinal in nature. The analysis of these data is to compare covariate sub-groups and draw region- and family-level inferences in terms of SRH. Parameters and random-effects are sampled from the joint posterior densities following a Bayesian paradigm. Three criteria are used for model selection: Watenable information criterion, log pseudo marginal likelihood, and deviance information criterion. All three suggest that we need to account for both region- and family-level variabilities in order to model SRH. The extent to which the models replicate the observed data is examined by posterior predictive checks. Differences in SRH are found between levels of economic and demographic variables, regions of Turkey, and families who participated in the survey. Some of the interesting findings are that unemployed people are 19% more likely to report poorer health than employed people, and rural Aegean is the region that has the least probability of reporting poorer health.

***Öz***

***Türkiye istatistiki bölge ve aile düzeyinde kümelenmiş sıralı sonuç verisinin Bayesçi modellemesi***

*Bu çalışma, üç seviyeli sıralı sonuç verisinin, rastgele etkili terimler içeren polytomous lojistik regresyon ile analizi üzerinedir. Rastgele etkili terimlerin, regresyon katsayıları için marjinal yorumlar elde edilebilmesini mümkün kılan logit linki için Bridge dağılımını takip ettikleri varsayılmıştır. Veri Türkiye Gelir ve Yaşam Koşulları Çalışması'ndan elde edilmiştir. Sonuç değişkeni sıralı bir yapıya sahip olan algılanan sağlık düzeyidir (ASD). Bu verinin analizi ile, bağımsız değişenlerin alt grupları, bölge ve aile düzeyinde ASD hakkında çıkarımlar yapılması amaçlanmaktadır. Bayesçi paradigma takip edilerek parametre ve rastgele etkilerin bileşik sonsal dağılımından örnekler elde edilmiştir. Model seçimi için üç kriter kullanılmıştır: Watanebe bilgi kriteri, log yalancı marjinal olabilirlik, ve sapma bilgi kriteri. Üç kriter de, bölge ve aile düzeyindeki varyasyonların, algılanan sağlık düzeyinin modellenmesi için göz önünde bulundurulması gerektiğine işaret etmektedir. Modellerin, gözlenen veriye benzer veriler üretme yeterliliğini anlamak için sonsal kestirim kontrolleri yapılmıştır. Ekonomik ve demografik değişkenlerin seviyeleri, Türkiye'nin bölgeleri ve çalışmaya dahil edilen aileler arasında ASD açısından farklılıklar bulunmuştur. Örneğin, işsiz insanlar çalışan insanlara kıyasla %19 daha yüksek ihtimalle kötü sağlık durumu raporlarken, kırsal Ege kötü sağlık durumu raporlama konusunda en düşük olasılığa sahip bölgedir.*

## 1. Introduction

In this study, we consider the analysis of three-level ordinal outcome data. The data come from the Turkish Income and Living Conditions Surveys (TR-SILC) conducted by the Turkish Statistical Institute since 2006. In TR-SILC, the data are collected as panels of four years and cross-sectionally. Since regional information is only available in the cross-sectional data, in this study we consider the cross-section of one year; for three-level analysis of panel data, interested reader is referred to [1].

In the cross-sections of TR-SILC, data are collected on individuals that are nested within families. One would expect individuals from the same family to be more similar compared to individuals from other families, e.g. due to genetic factors, lifestyle, economic conditions, etc. The data is further nested within the statistical regions of Turkey. There are 12 statistical regions, and in addition, we have the information about rural and urban areas. Thus, there are 24 regional units in total. It is expected that individuals from the same region are more similar than those from other regions.

The outcome of interest is self-rated health (SRH), which can take one of the following values: very poor, poor, fair, good, very good. A number of family and individual level explanatory variables are available. The main research interest of this study is to understand:

- the relationships between SRH and explanatory variables, and
- the region- and family-specific characteristics.

To address these, we consider a polytomous logistic regression model with random-effects. The presence of random-effects in a regression framework makes the interpretation of the regression coefficients, i.e. the first research interest, conditional on two persons from different covariate groups having the same random-effect. This is a restrictive assumption, as one would typically expect the random-effects associated with these two persons to be different. Following [1] and [2], and the references therein, we assume that the random-effects have a Bridge distribution for the logit link [3]. This assumption allows for an unconditional (or marginal) interpretation of the regression coefficients as in the classical regression setting (without random-effects). We take a Bayesian paradigm, and sample the parameters and random-effects from the joint posterior densities using Hamiltonian Monte Carlo (HMC, [4]).

The rest of the paper is organised as follows. In Section 2, we present the 2013 cross-section of TR-SILC. In Section 3, we present the modelling framework and the model selection criteria. Section 4 presents the results, while Section 5 the posterior predictive checks. Section 6 closes the paper with conclusion and discussion.

**2. Data**

The Turkish Income and Living Conditions Study (TR-SILC) surveys collect detailed information on income, poverty, social exclusion, living conditions, housing, labour, education and health. Turkey has been conducting the survey since 2006 as part of its integration into the EU, in the form of 4-year panels and cross-sectional surveys. For the details of TR-SILC and SILC in general, the interested reader is referred to [1] and [5] and the references therein.

In this study, we consider a cross-section (specifically, the 2013 data) to examine, in particular, regional differences in health, as regional information is not available in the panels. The outcome variable is self-rated health (SRH) which is ordinal and can take one of the following values: very poor, poor, fair, good, very good. SRH represents the general health status of an individual and is considered as a predictor of morbidity and mortality [6]. Following [7] and [8], we consider a re-categorised version of the variable as good health (good/very good), fair health and poor health (poor/very poor). Mean household disposable income, defined as total family income divided by family size (MHDI, in Turkish Lira), gender (male, female), marital status (married, never married, other), age (15 - 34, 35 - 64, 65+), education level (primary school or less, secondary or high school, higher education), working status (full/part time work, unemployed, student, housekeeping, other) are the explanatory variables. Note that, MHDI is a family-level variable, while the other variables are at individual-level.

The 2013 cross-section includes 53,496 individuals from 19,899 families. Summary statistics for the variables can be found in Table 1. The SRH distribution with respect to regions is depicted in Figure 1. In the analyses, the MDHI will be used in natural logarithm scale, because the variable is right-skewed. Since there are only 74 individuals from rural Istanbul, the data from rural and urban Istanbul are combined in the analyses (and simply referred to as Istanbul). There is no missing data in the variables considered.

Table 1. Summary statistics for the 2013 cross-section of the TR-SILC data.

|  | **Poor** | **Fair** | **Good** |
|---|---|---|---|
| *Mean household disposable income* | | | |
| Minimum | 375.7 | 44.2 | 6.3 |
| 25th percentile | 4,125.0 | 4,991.0 | 5,186.4 |
| Median | 6,316.0 | 7,550.0 | 8,057.2 |
| Mean | 7,494.8 | 9,532.3 | 10,934.1 |
| 75th percentile | 9,186.4 | 11,300.0 | 12,663.4 |
| Maximum | 178,842.3 | 210,667.3 | 373,924.6 |
| Standard deviation | 5,862.5 | 8,832.9 | 11,175.3 |
| # of individuals | 7,162 | 11,280 | 35,054 |
| *Family size* | | | |
| Minimum | 1.0 | 1.0 | 1.0 |
| 25th percentile | 2.0 | 2.0 | 2.0 |
| Median | 3.0 | 3.0 | 3.0 |
| Mean | 3.2 | 3.1 | 3.4 |
| 75th percentile | 4.0 | 4.0 | 4.0 |
| Maximum | 13.0 | 17.0 | 17.0 |
| Standard deviation | 1.6 | 1.5 | 1.6 |
| # of individuals | 7,162 | 11,280 | 35,054 |
| *Gender* | | | |
| Female | 4,377 (15.8%) | 6,436 (23.3%) | 16,820 (60.9%) |
| Male | 2,785 (10.8%) | 4,844 (18.7\%) | 18,234 (70.5%) |
| *Marital Status* | | | |
| Married | 4,753 (13.2%) | 8,721 (24.2%) | 22,521 (62.6%) |
| Never married | 660 (5.2%) | 883 (6.9%) | 11,168 (87.9%) |
| Other | 1,749 (36.5%) | 1,676 (35.0%) | 1,365 (28.5%) |
| *Age* | | | |
| 15-34 | 844 (3.8%) | 1,986 (9.0%) | 19,342 (87.2%) |
| 35-64 | 3,602 (14.3%) | 6,998 (27.7%) | 14,641 (58.0%) |
| 65+ | 2,716 (44.6%) | 2,296 (37.7%) | 1,071 (17.6%) |
| *Education level* | | | |
| Primary school or less | 6,235 (21.4%) | 8,438 (29.0%) | 14,406 (49.5%) |
| Secondary or high school | 807 (4.3%) | 2,207 (11.7%) | 15,830 (84.0%) |
| Higher education | 120 (2.2%) | 635 (11.4%) | 4,818 (86.5%) |
| *Working status* | | | |
| Full/part time workers | 1,550 (6.3%) | 4,558 (18.6%) | 18,414 (75.1%) |
| Unemployed | 127 (6.2%) | 305 (15.0%) | 1,606 (78.8%) |
| Housekeeper | 1,945 (13.7%) | 3,696 (26.1%) | 8,534 (60.2%) |
| Retired | 961 (22.1%) | 1,556 (35.8%) | 1,835 (42.2%) |
| Student | 65 (1.5%) | 183 (4.2%) | 4,102 (94.3%) |
| Other | 2,514 (61.9%) | 982 (24.2%) | 563 (13.9%) |

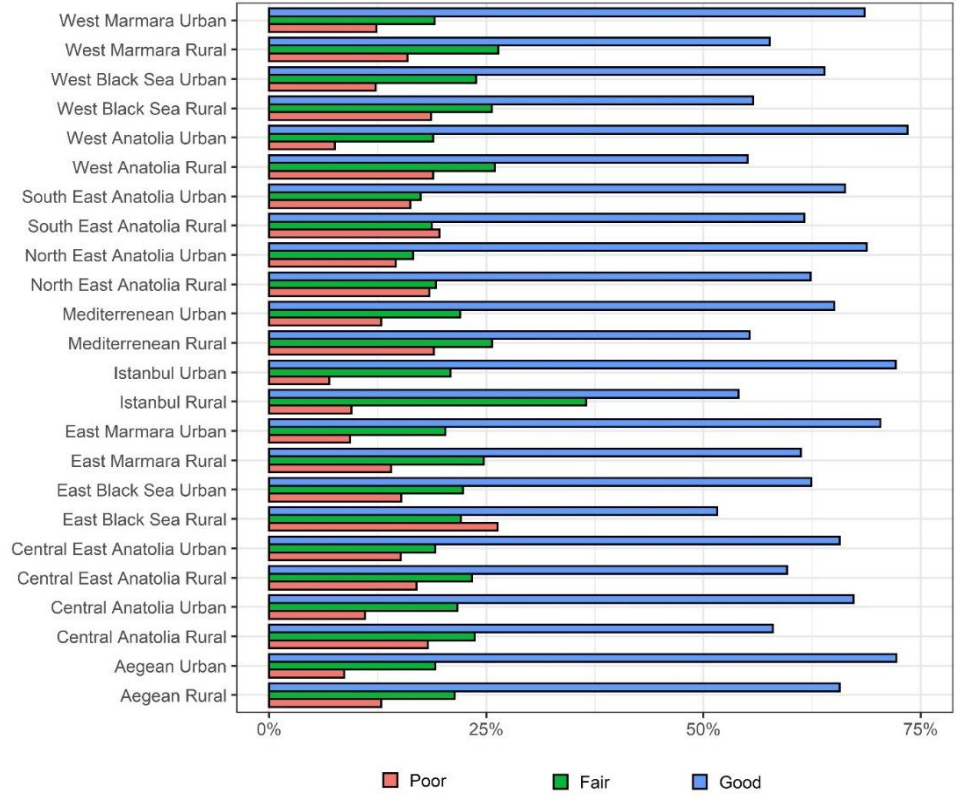

Figure 1. SRH distributions with respect to the statistical regions.

## 2. Modelling framework

### 2.1. Notation and model

Let $Y_{ijk} \in \{1 = \text{good health}, 2 = \text{fair health}, 3 = \text{poor health}\}$ be the outcome belonging to individual $k$ $(k = 1, \ldots, n_{ij})$ from family $j$ $(j = 1, \ldots, m_i)$ and region $i$ $(i = 1, \ldots, s)$. Also let $x_{ijk}$ a $p \times 1$ dimensional covariate matrix.

The modelling framework that we consider to understand the relationships between SRH and the explanatory variables whilst taking into account the region- and family-level variabilities has the following form:

$$\text{logit}\{P(Y_{ijk} \le a | x_{ijk}, U_i, V_{ij}, \theta)\} = \alpha_a^c - x_{ijk}^T \beta^c - U_i - V_{ij}, \ a = 1,2, \tag{1}$$

where in addition to the notation introduced before, $\text{logit}(x) = \log\left(\frac{x}{1-x}\right)$, $P(\cdot)$ the probability operator, $\alpha_a^c$ category-specific threshold parameters, $\beta^c$ regression coefficients, $U_i$ and $V_{ij}$ are random-effects, and $\theta$ a generic notation for parameters. In this setting, the interpretations of $\alpha_a^c$ and $\beta^c$ are conditional on the $U_i$ and $V_{ij}$ terms being the same for two persons belonging to two different covariate groups; the super-script "c" stands for conditional interpretation. Assuming the random-effects having a Bridge distribution for the logit link allows us to directly obtain the unconditional/marginal interpretation, i.e. as in the usual regression setting. We call these parameters as the marginal parameters and denote by $\alpha_a^m$ and $\beta^m$.

*2.2. Bridge distributed random-effects*

One can obtain the relationships between $\alpha_a^m$ and $\alpha_a^c$, and $\beta^m$ and $\beta^c$ by solving the following equation:

$$P(Y_{ijk} \leq a | x_{ijk}, \alpha_a^m, \beta^m) = E_{U,V}\left(P(Y_{ijk} \leq a | x_{ijk}, U_i, V_{ij}, \alpha_a^c, \beta^c, \theta_{U,V})\right), \tag{2}$$

where $E(\cdot)$ is the expectation operator, $\theta_{U,V}$ are the parameters of $U_i, V_{ij}$. The relationships would be available in closed-form, when one assumes Bridge distribution for the random-effects, as follows. Let $U_i = U_i^*/\emptyset_V$, where $[U_i^*] = \text{Bridge}(\emptyset_{U^*})$, and $[V_{ij}] = \text{Bridge}(\emptyset_V)$, with $0 < \emptyset_{U^*}, \emptyset_V < 1$, and "[·]" denotes "the distribution of". One can then obtain the marginal estimates as $\alpha_a^m = \emptyset_{U^*}\emptyset_V \alpha_a^c$ and $\beta^m = \emptyset_{U^*}\emptyset_V \beta^c$, see [1].

Under the above specification, note that $U_i$ is no longer Bridge-distributed, but it has a Modified Bridge distribution. Properties of the Bridge and Modified Bridge distributions are presented below.

The probability density function of the Bridge distribution for logit link [3] is given by

$$f(x|\emptyset) = \frac{1}{2\pi}\frac{\sin(\emptyset\pi)}{\cosh(\emptyset x)+\cos(\emptyset\pi)}, -\infty < x < \infty, 0 < \emptyset < 1. \tag{3}$$

where $\cosh(\cdot)$ is the hyperbolic cosine, defined as $\cosh(x) = \frac{1}{2}(\exp(x) + \exp(-x))$. It is a symmetric distribution, has zero-mean and a variance of $\frac{\pi^2}{3}(\emptyset^{-2} - 1)$. The density function of the modified Bridge distribution, for generic $X$, $Y$ and $Z$ with $X = Y/\emptyset_Z$, $[Y|c] = \text{Bridge}(\emptyset_Y)$, $[Z|\emptyset_Z] = \text{Bridge}(\emptyset_Z)$, is given by

$$f(x|\emptyset_Y, \emptyset_Z) = \frac{\emptyset_Z}{2\pi}\frac{\sin(\emptyset_Y\pi)}{\cosh(\emptyset_Y\emptyset_Z x)+\cos(\emptyset_Y\pi)}, -\infty < x < \infty, 0 < \emptyset_Y, \emptyset_Z < 1, \tag{4}$$

Modified Bridge is also symmetric, zero-mean, and has a variance of $\frac{\pi^2}{3\emptyset_Z^2}(\emptyset_Y^{-2} - 1)$.

*2.3. Priors and inference*

We select weakly informative prior distributions for the parameters following the literature. For $\alpha_a^c$ and $\beta^c$, Cauchy distribution with location parameter 0 and scale parameter 5 is considered [9]. For Bridge distribution, the standard deviation, $\frac{\pi^2}{3}(\emptyset^{-2} - 1)$, is assumed to be half-Cauchy with location 0 and scale 5 [10, 11].

We sample the parameters and the random-effects from the joint posterior densities using the No-U-Turn Sampler [12], which is a modified version of Hamiltonian Monte Carlo [4]. Details of the posterior distributions are skipped here; for details one can consult the work of [1]. For computation, we use the R [13] package mixed3 (https://github.com/ozgurasarstat/mixed3).

*2.4. Model selection*

For model selection, we consider three widely used criteria that are used within the Bayesian framework. First of these is the Watanebe Information Criterion (WIC, [14]):

$$\text{WAIC} = -2(\text{lppd} - \rho), \tag{5}$$

where, "lppd" stands for log point-wise posterior density that is calculated as

$$\text{lppd} = \sum_{i=1}^{S} \sum_{j=1}^{m_i} \sum_{k=1}^{n_{ij}} \log\left(\frac{1}{M} \sum_{l=1}^{M} [Y_{ijk} | U_i^{(l)}, V_{ij}^{(l)}, \theta^{(l)}]\right), \tag{6}$$

and $\rho$ is the effective number of parameters and calculated as

$$\rho = \sum_{i=1}^{S} \sum_{j=1}^{m_i} \sum_{k=1}^{n_{ij}} V_{l=1}^{M}\left(\log\left([Y_{ijk} | U_i^{(l)}, V_{ij}^{(l)}, \theta^{(l)}]\right)\right), \tag{7}$$

with

$$V_{l=1}^{M}(a) = \frac{1}{M}\sum_{l=1}^{M}(a^{(l)} - \bar{a})^2. \tag{8}$$

In (6-8), the superscript $(l)$ denotes the $l$th draw of the associated term from the joint posterior densities, $M$ the size of the HMC sample. Note that lower values of WAIC indicate better model performance.

The second is the log pseudo marginal likelihood (LPML, [15, 16]). It is calculated as

$$\text{LMPL} = \sum_{i=1}^{S} \sum_{j=1}^{m_i} \sum_{k=1}^{n_{ij}} \log\left(\widehat{CPO_{ijk}}\right), \tag{9}$$

where CPO stands for conditional predictive ordinate that is defined as leave-one-out cross-validated predictive density, $CPO_{ijk} = [Y_{ijk} | Y_{-(ijk)}]$, where $Y_{-(ijk)}$ denotes the full set of outcomes without the observation $(ijk)$. The estimate of CPO that we use is the harmonic mean estimate [15],

$$\widehat{CPO_{ijk}} = \left(\frac{1}{M}\sum_{l=1}^{M} \frac{1}{[Y_{ijk} | U_i^{(l)}, V_{ij}^{(l)}, \theta^{(l)}]}\right)^{-1}. \tag{10}$$

Larger values of LMPL indicate better model fit.

The third criterion is the deviance information criterion (DIC, [17]) for which the formula is given by

$$DIC = 2\bar{D} - D(\bar{\theta}, \bar{U}, \bar{V}), \tag{11}$$

where

$$\bar{D} = \frac{1}{M}\sum_{l=1}^{M} -2\log\left([Y_{ijk} | U_i^{(l)}, V_{ij}^{(l)}, \theta^{(l)}]\right), \tag{13}$$

$$D(\bar{\theta}, \bar{U}, \bar{V}) = -2\log\left(\sum_{i=1}^{S}\sum_{j=1}^{m_i}\sum_{k=1}^{n_{ij}} [Y_{ijk} | \bar{\theta}, \bar{U}_i, \bar{V}_{ij}]\right), \tag{14}$$

and $\bar{\theta} = \frac{1}{M}\sum_{l=1}^{M} \theta^{(l)}$, $\bar{U}_i = \frac{1}{M}\sum_{l=1}^{M} U_i^{(l)}$, $\bar{V}_{ij} = \frac{1}{M}\sum_{l=1}^{M} V_{ij}^{(l)}$. Lower values of DIC indicate better fit.

## 4. Results

We fit the following three models to the 2013 cross-section of the TR-SILC:

- fixed-effects: no $U_i$ and $U_{ij}$ terms in model (1),
- two-level: no $U_i$ term in model (1),
- three-level: the model described in (1).

For each model, we run 4 parallel HMC chains started from random initials. Each chain has the length of 2,000, first halves of which are discarded as the burn-in. In total, the HMC chains have the size of 4,000. To assess the convergence of the chains, we use trace-plots, density plots, and the R-hat statistic [18]. It took about 1.8, 8.6, and 8.8 hours to fit the fixed-effects, two-level and three-level models, respectively, on a 64-bit personal laptop with 12 GB RAM and Intel(R) Core(TM) i5-8250U CPU @ 1.60GHz running Windows 10. Means, standard deviations (sd) and 2.5%th and 97.5%th percentiles of the HMC samples are presented in Table 2. For the two- and three-level models, we directly present the $\alpha_a^m$ and $\beta^m$, as $\alpha_a^c$ and $\beta^c$ are not of primary interest. The model selection criteria are presented in Table 3. All of the LPML, WAIC and DIC indicate that the three-level model is the best fitting model, whereas the fixed-effects model is the worst. This indicates that both the regional- and family-level dependencies need to be taken into account in order to appropriately analyse the TR-SILC data.

Since the three-level model is found to be the best fitting model, here we only interpret the related coefficients. One percent decrease in MDHI was associated with approximately 3% ($= (\exp(0.293 \times \log(1.1)) - 1) \times 100$) increase in the odds of reporting poorer health. Females were approximately 28% ($= (\exp(0.244) - 1) \times 100$) more likely to report poorer health compared to males. People who never married were approximately 33% less likely to report poorer health compared to people who were married, whereas people whose marital status was different than married/never married were approximately 30% more likely to report poorer health compared to those who were married. People whose age was in the $35 - 64$ and $65 +$ categories were 2.3 and 5.7 times more likely to report poorer health compared to those who were in the $15 - 34$ category, respectively. As the education level increased the probability of reporting poorer health decreased. Students were less likely to report poorer health compared to employed people. People in all the other working status categories were more likely to report poorer health. For example, unemployed people were 19% more likely to report poorer heath compared to those were employed. Means, and 2.5%th and 97.5%th percentiles of the HMC samples of the $U_i$ terms are displayed in Figure 2. Rural and urban Aegean regions are the ones with the lowest chance of reporting poorer health. Urban and rural East and West Marmara regions, Istanbul and urban West Anatolia are also amongst the lowest risk regions. Rural and urban East Black Sea regions are the ones that had the highest chance of reporting poorer health. Both urban and rural Central East Anatolia are also amongst the regions that had the highest chance. Means, 2.5%th and 97.5%th percentiles of the HMC samples of the $V_{ij}$ terms for randomly selected 50 families are displayed in Figure 3. Two- and three-level models largely agree on the mean estimates, whereas we see minor differences in the 95% credibility intervals.

Table 2. Estimation results. "sd" stands for standard deviation.

| | | Fixed-effects | | | Two-level | | | Three-level | | |
|---|---|---|---|---|---|---|---|---|---|---|
| **Variable** | **Parameter** | **Mean** | **sd** | **2.5%, 97.5%** | **Mean** | **sd** | **2.5%, 97.5%** | **Mean** | **sd** | **2.5%, 97.5%** |
| Threshold | $\alpha_1^m$ | -0.557 | 0.169 | -0.894, -0.232 | -0.835 | 0.186 | -1.198, -0.470 | -0.229 | 0.194 | -0.607, 0.162 |
| Threshold | $\alpha_2^m$ | 1.126 | 0.168 | 0.793, 1.444 | 0.821 | 0.186 | 0.460, 1.186 | 1.371 | 0.192 | 0.997, 1.757 |
| log(MHDI) | $\beta_1^m$ | -0.351 | 0.017 | -0.385, -0.319 | -0.377 | 0.019 | -0.413, -0.340 | -0.293 | 0.020 | -0.333, -0.253 |
| Male (Ref) | - | - | - | - | - | - | - | - | - | - |
| Female | $\beta_2^m$ | 0.250 | 0.027 | 0.196, 0.303 | 0.247 | 0.025 | 0.200, 0.297 | 0.244 | 0.024 | 0.196, 0.291 |
| Married (Ref) | - | - | - | - | - | - | - | - | - | - |
| Never married | $\beta_3^m$ | -0.288 | 0.039 | -0.366, -0.212 | -0.297 | 0.040 | -0.374, -0.220 | -0.282 | 0.039 | -0.358, -0.205 |
| Other | $\beta_4^m$ | 0.255 | 0.035 | 0.188, 0.326 | 0.265 | 0.034 | 0.198, 0.333 | 0.263 | 0.033 | 0.198, 0.329 |
| 15-34 (Ref) | - | - | - | - | - | - | - | - | - | - |
| 35-64 | $\beta_5^m$ | 1.194 | 0.030 | 1.137, 1.253 | 1.218 | 0.030 | 1.158, 1.277 | 1.186 | 0.032 | 1.122, 1.251 |
| 65+ | $\beta_6^m$ | 1.963 | 0.042 | 1.881, 2.048 | 1.960 | 0.042 | 1.874, 2.041 | 1.906 | 0.049 | 1.810, 2.002 |
| Higher education (Ref) | - | - | - | - | - | - | - | - | - | - |
| Primary or less | $\beta_7^m$ | 0.925 | 0.045 | 0.834, 1.015 | 0.861 | 0.046 | 0.772, 0.952 | 0.843 | 0.046 | 0.751, 0.931 |
| Secondary or high school | $\beta_8^m$ | 0.292 | 0.046 | 0.202, 0.384 | 0.283 | 0.046 | 0.190, 0.373 | 0.286 | 0.045 | 0.198, 0.377 |
| Full/part time (Ref) | - | - | - | - | - | - | - | - | - | - |
| Housekeeper | $\beta_9^m$ | 0.278 | 0.030 | 0.220, 0.336 | 0.280 | 0.029 | 0.224, 0.337 | 0.264 | 0.028 | 0.207, 0.321 |
| Other | $\beta_{10}^m$ | 2.131 | 0.043 | 2.047, 2.212 | 2.108 | 0.042 | 2.027, 2.192 | 2.020 | 0.049 | 1.921, 2.114 |
| Retired | $\beta_{11}^m$ | 0.785 | 0.035 | 0.716, 0.853 | 0.730 | 0.035 | 0.662, 0.799 | 0.724 | 0.034 | 0.658, 0.791 |
| Student | $\beta_{12}^m$ | -0.458 | 0.076 | -0.608, -0.311 | -0.294 | 0.070 | -0.435, -0.160 | -0.284 | 0.066 | -0.411, -0.155 |
| Unemployed | $\beta_{13}^m$ | 0.202 | 0.062 | 0.082, 0.320 | 0.177 | 0.057 | 0.062, 0.288 | 0.173 | 0.058 | 0.060, 0.284 |
| $U^*$ | $\phi_{U^*}$ | - | - | - | 0.816 | 0.006 | 0.805, 0.827 | 0.959 | 0.013 | 0.930, 0.979 |
| $V$ | $\phi_V$ | - | - | - | - | - | - | 0.821 | 0.006 | 0.810, 0.832 |

Table 3. Model selection results

| Model | LPML ↑ | WAIC ↓ | DIC ↓ |
|---|---|---|---|
| Fixed-effects | -37,267.8 | 74,535.5 | 74,535.3 |
| Two-level | -35,919.6 | 71,315.8 | 71,781.5 |
| Three-level | -35,830.2 | 71,158.0 | 71,603.4 |

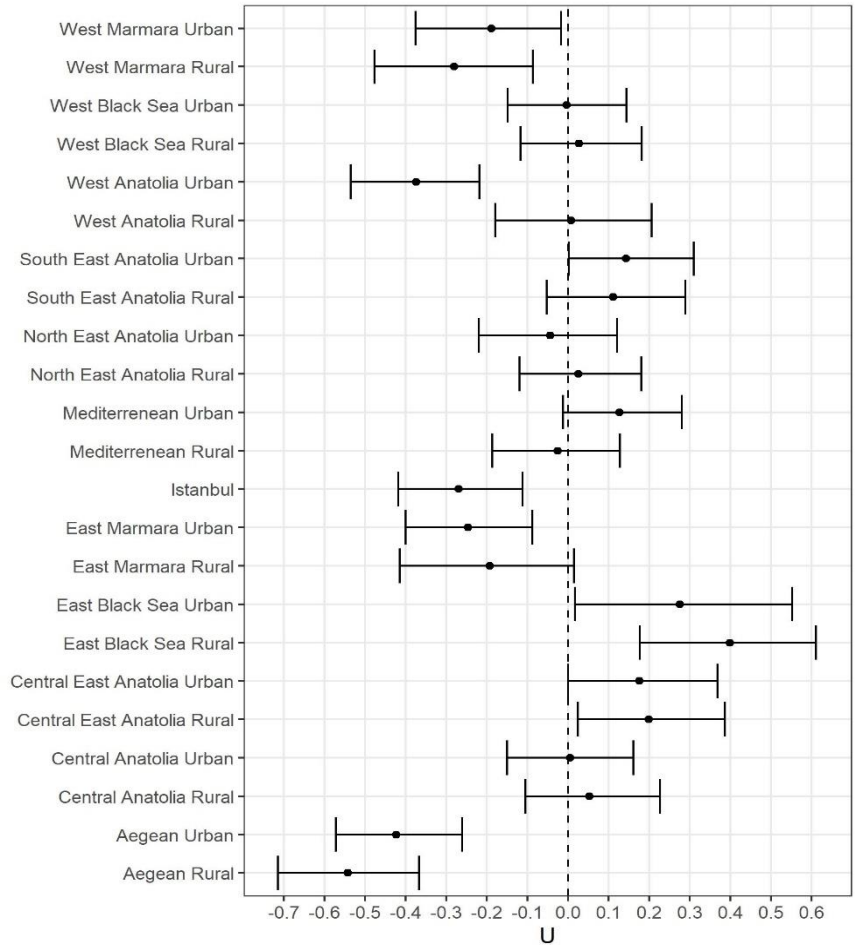

Figure 2. Means (in black dots) and 2.5%th and 97.5%th percentiles (as error bars) of the posterior distributions of the $U$ terms based on the three-level model.

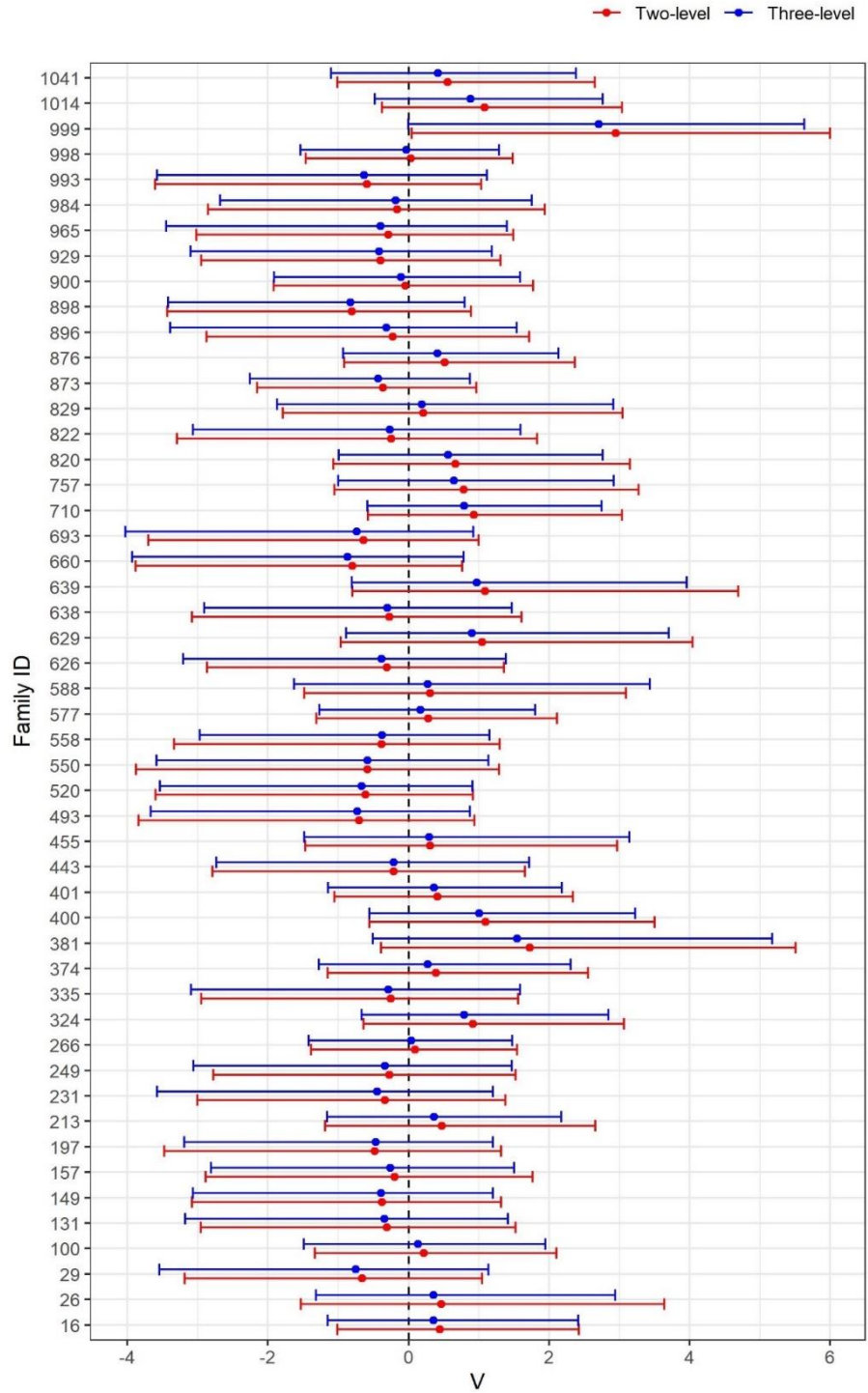

Figure 3. Means (in dots) and 2.5%th and 97.5%th percentiles of the HMC samples of the family-level random-effects ($V_{ij}$) for randomly selected 50 families, based on the two-level model (in red) and three-level model (in blue).

## 5. Posterior predictive checks

In order to see how the fitted models replicate the SRH outcome data, we performed posterior predictive checks. We simulated data for each of the 4,000 elements of the HMC samples from

$$[Y_{ijk}^{sim}|Y] = \int \int \int [Y_{ijk}^{sim}|U_i, V_{ij}, \theta][U_i|Y, \theta][V_{ij}|Y, \theta][\theta|Y] dU\, dV\, d\theta \qquad (14)$$

for the three-level model, from

$$[Y_{ijk}^{sim}|Y] = \int \int [Y_{ijk}^{sim}|V_{ij}, \theta][V_{ij}|Y, \theta][\theta|Y]\, dV\, d\theta \qquad (15)$$

for the two-level model, and from

$$[Y_{ijk}^{sim}|Y] = \int [Y_{ijk}^{sim}|\theta][\theta|Y]\, d\theta \qquad (16)$$

for the fixed-effects model, Y indicates the set of observed SRH outcomes. We then compared the simulated data-sets with the observed SRH outcomes. We report means, standard deviations and 2.5%th and 97.5%th percentiles for the percentages of matches and mis-matches between the observed and simulated SRH outcomes, see Table 4. Here, matches and mis-matches are defined as

- ``-2'': observed outcome being ``good health'' and simulated being ``poor health'';
- ``-1'': observed being ``good health'' and simulated being ``fair health'', or observed being ``fair health'' and replicated being ``poor health'';
- ``0'': observed and simulated being the same;
- ``1'': observed being ``fair health'' and simulated being ``good health'', or observed being ``poor health'' and simulated being ``fair health'';
- ``2'': observed being ``poor health'' and replicated being ``good health''.

Note that non-zero values mean mis-match, whereas ``-2'' and ``2'' would mean the most mis-match. Two- and three-level models seem to perform similarly in terms of replicating the observed data, whereas fixed-effects model seems to be the worst.

Table 4. Posterior predictive check results. ``Diff'' stands for difference, ``sd'' for standard deviation.

|      | **Fixed-effects** | | | **Two-level** | | | **Three-level** | | |
|------|------|------|------|------|------|------|------|------|------|
| **Diff** | **Mean** | **sd** | **2.5%, 97.5%** | **Mean** | **sd** | **2.5%, 97.5%** | **Mean** | **sd** | **2.5%, 97.5%** |
| -2 | 8.63 | 0.13 | 8.37, 8.89 | 6.54 | 0.12 | 6.32, 6.78 | 7.10 | 0.26 | 6.59, 7.60 |
| -1 | 16.69 | 0.18 | 16.34, 17.03 | 14.46 | 0.17 | 14.12, 14.81 | 15.44 | 0.37 | 14.69, 16.13 |
| 0 | 49.30 | 0.19 | 48.93, 49.66 | 51.95 | 0.17 | 51.61, 52.29 | 51.04 | 0.39 | 50.30, 51.82 |
| 1 | 16.58 | 0.10 | 16.38, 16.78 | 17.49 | 0.09 | 17.31, 17.66 | 17.14 | 0.14 | 16.86, 17.42 |
| 2 | 8.80 | 0.07 | 8.67, 8.93 | 9.56 | 0.06 | 9.43, 9.68 | 9.29 | 0.12 | 9.06, 9.51 |

## 5. Conclusion and discussion

In this study, we analysed the 2013 cross-section of the TR-SILC study. The outcome variable is the SRH which has three categories: poor, fair and good health. A number of economic and demographic variables are considered to explain the variability in SRH. The data has two sources of dependency: statistical regions and families. We considered a polytomous logistic regression with Bridge distributed random-effects. The Bridge distribution specifically allows us to obtain marginal interpretations of the regression coefficients, while making inferences at the region- and family-level. Inferences for parameters and random-effects are obtained under the Bayesian paradigm. The methods are implemented in the R package mixed3.

We found differences between covariate subgroups with respect to SRH. People with higher income and education were less likely to report poorer health overall. Gender, marital status, and age also appear to explain variability in SRH. People who have never been married appear less likely to have poorer health. Similarly, students seem to be less likely to report poorer health compared to those who are employed. We shall note that both of these results can be explained by the age factor.

It is interesting to observe differences between regions in terms of reporting poorer health. The Aegean and Marmara regions have the lowest probability of reporting poorer health, while East Black Sea and Central East Anatolia have the highest probability of reporting poorer health. It is also interesting to observe differences between the families through the random-effects, which can be considered as proxies for unmeasured characteristics, e.g. genetic factors. Besides these observations, the model selection criteria we considered suggest that both regional- and family-level dependencies need to be taken into account when analysing the TR-SILC data.

This paper is the first to consider appropriate statistical modelling for the analysis of cross-sections of TR-SILC, where we analysed data from the 2013 cross-section. Other cross-sections can also be analysed and the results are compared. Causal inference can be considered to draw causal interpretations, as the TR-SILC data is observational. These are the beyond the scope of this work.

**Acknowledgements**

The author acknowledges the encouragements of and helpful discussions with Dr. Kutsev Bengisu Özyörük while writing this paper. The author also thanks to Dr. Mahmut Yardım for introducing the SILC studies and helpful discussions.